\documentclass[aps,prx,twocolumn,superscriptaddress,nobibnotes]{revtex4-2}

\usepackage{graphicx}
\usepackage{dcolumn}
\usepackage{bm}
\usepackage{amsmath}
\usepackage{amssymb}
\usepackage{latexsym}
\usepackage{epsfig}
\usepackage{amsbsy}
\usepackage{array}
\usepackage{amssymb}
\usepackage{setspace}

\usepackage[justification=centerlast]{caption}
\usepackage{subfig}
\usepackage{booktabs}

\begin{document}
	
\title{Quantum Evolution Time Limit: a General Non-Markovian Bound}

\author{Xiangyi~Meng}%
\email{xm@bu.edu}
\affiliation{State Key Laboratory of Advanced Optical Communication$\text{,}$ Systems and Networks$\text{,}$ Department of Electronics$\text{,}$ and Center for Quantum Information Technology, Peking University, Beijing 100871, China}%
\affiliation{Department of Physics, Boston University, Boston, Massachusetts 02215, USA}%
\author{Chengjun~Wu}%
\affiliation{State Key Laboratory of Advanced Optical Communication$\text{,}$ Systems and Networks$\text{,}$ Department of Electronics$\text{,}$ and Center for Quantum Information Technology, Peking University, Beijing 100871, China}
\author{Hong~Guo}%
\email{hongguo@pku.edu.cn}
\affiliation{State Key Laboratory of Advanced Optical Communication$\text{,}$ Systems and Networks$\text{,}$ Department of Electronics$\text{,}$ and Center for Quantum Information Technology, Peking University, Beijing 100871, China}

\date{\today}

\begin{abstract}
We derive a sharp bound as the quantum speed limit (QSL) for the minimal evolution time of quantum open systems in the non-Markovian strong-coupling regime with initial mixed states by considering the effects of both renormalized Hamiltonian and dissipator. For a non-Markovian quantum open system, the possible evolution time between two arbitrary states is not unique, among the set of which we find that the minimal one and its QSL can decrease more steeply by adjusting the coupling strength of the dissipator, which thus provides potential improvements of efficiency in many quantum physics and quantum information areas.
\end{abstract}

\maketitle

\section*{Introduction}
As a fundamental bound for the evolution time of quantum systems, the quantum speed limit (QSL) (also referred to as quantum evolution time limit) plays an important role in tremendous areas of quantum physics and quantum information, such as quantum computation and communication~\cite{WorksComputation, WorksCommun}, quantum metrology~\cite{WorksMetrology}, cavity quantum electrodynamics~\cite{WorksQED1}, quantum control~\cite{QuantumSpeedUp}, etc. The derivation of QSL is most required for the purpose of simplification and/or optimization in theoretical analysis, since in most quantum-cases one only needs to derive a lower bound on the minimal time of evolution without solving the exact equation to see the dominant factors in evolution and/or optimize our demand. For closed quantum systems, two types of QSL have been derived at the start: the Mandelstam-Tamm (MT) bound $\tau \ge \pi \hbar /\left( 2\Delta E \right)$~\cite{MT} and the Margolus-Levitin (ML) bound $\tau \ge \pi \hbar /\left( 2\left\langle E \right\rangle  \right)$~\cite{ML}. Since then, further investigations are launched into QSL~\cite{QSL8,QSL9,QSL5,QSL3,QSL2}. As the energy of a closed system is conserved, the QSL of a closed system is decided by the variance of energy $\Delta E$ or the mean energy $\left\langle E \right\rangle$, related only to the unitary Hamiltonian. Recently, the QSL for quantum open systems~\cite{OpenQuantumSystems} draws wide attention with several bounds~\cite{QSLopen2,QSLopen1,QSLnonmarkov,QSLnonmarkov2,QSL10} being found. Because there is energy and/or coherence exchange between system and environment for quantum open systems, the evolution generator therein contains not only a time-dependent Hamiltonian $H_{t}$ but also a dissipator ${{\mathcal{D}}_{t}}\left( {{\rho }_{t}} \right)$ (a trace-preserving term referring to dissipation behaviors)~\cite{OpenQuantumSystems}. In quantum open systems, non-Markovianity is valuable in practice and highly emphasized for its particular characteristics of memory effect, negative energy/population flow and singularity of the state evolution~\cite{NMarkov2, NMarkov3}. The latter two characteristics are commonly found in the strong-coupling regime, where the system and environment are strongly coupled and the non-Markovianity becomes a non-negligible strong effect~\cite{NMarkov1}. Typically, a strong-coupling regime can be achieved and temporarily maintained in high-$Q$ optical micro-cavities~\cite{WorksCavity2} and quantum circuits~\cite{WorksCircuits}. In spite of recent breakthrough on measurement methods for non-Markovianity~\cite{DegreeOfNonMarkovianity1, DegreeOfNonMarkovianity3, AssessNonMarkovianity, MeasureOfNonMarkovianity0,DegreeOfNonMarkovianitySummary}, the strong-coupling regime still remains as an open question. Also, QSL issue becomes more complicated than it was considered~\cite{QSLnonmarkov}, since in such a regime the possible evolution time between two arbitrary states is not unique, while only the QSL for the minimal one does matter. In addition to non-Markovianity, the evolution of mixed states in quantum open systems also attracts concern. It is therefore of great significance to derive a sharp bound on evolution time for general conditions, i.e., for mixed states in different non-Markovian coupling regimes.

In this report, we study the non-Markovian problem by using geometric methods and derive a sharp bound for the minimal evolution time for quantum open systems with initial mixed states. We define the minimal evolution time $\hat{\tau}$ for non-Markovian quantum open systems as the minimal possible evolution time between two arbitrary states before we study its relevant QSL using new mathematical inequality tools. A steeper decrease of QSL than previous result~\cite{QSLnonmarkov} caused by strong non-Markovianity is observed in the examples of two-level models, indicating that a much smaller evolution time can be achieved in the strong-coupling regime. It is implied that the evolution of quantum physical process and computation involving strong-coupling interactions can be more effective.

\section*{Results}
\subsection*{Geometric fidelity}
To quantify the geometric distance between two general quantum states, the Bures fidelity~\cite{BuresFidelity} ${{\mathcal{F}}_{B}}\left( {{\rho }_{1}},{{\rho }_{2}} \right)={{\left\| \sqrt{{{\rho }_{1}}}\sqrt{{{\rho }_{2}}} \right\|}_{\operatorname{tr}}}$ with the Bures angle ${{\Theta }_{B}}=\cos^{-1} {{\mathcal{F}}_{B}}$ was usually used, where $\rho$ is the density operator of a general quantum state. Here, however, we introduce the \textit{relative-purity fidelity} ${{\mathcal{F}}_{R}}\left( {{\rho }_{1}},{{\rho }_{2}} \right)={{\left\| \sqrt{{{\rho }_{1}}}\sqrt{{{\rho }_{2}}} \right\|}_{\operatorname{HS}}}/\sqrt{{\left\| {{\rho }_{2}} \right\|}_{\operatorname{HS}}}$ with ${{\Theta }_{R}}=\cos^{-1} {{\mathcal{F}}_{R}}$. This one derived from the so-called relative purity~\cite{RelativeFidelity} is more useful in studying QSL~\cite{RelativeFidelityUSEFUL}. It is easy to prove that ${{\mathcal{F}}_{B}}\left( \rho ,\rho  \right)={{\mathcal{F}}_{R}}\left( \rho ,\rho  \right)=1$, and, if ${{\rho }_{2}}$ is a pure state, then one has ${{\mathcal{F}}_{B}}\left( {{\rho }_{1}},{{\rho }_{2}} \right)={{\mathcal{F}}_{R}}\left( {{\rho }_{1}},{{\rho }_{2}} \right)$.

From the von Neumann trace inequality~\cite{vonNeumann},
\begin{eqnarray*}
{\left\| \sqrt{{{\rho }_{1}}}\sqrt{{{\rho }_{2}}} \right\|}_{\text{HS}}^{2}
&=&\operatorname{Tr}\left\{ {{\rho }_{1}}{{\rho }_{2}} \right\}
\le
\sum\nolimits_{i}{\sigma _{1,i} \cdot \sigma _{2,i}}\\
&\le&
\sqrt{\sum\nolimits_{i}{\sigma _{1,i}^{2}}}\cdot \sqrt{\sum\nolimits_{i}{\sigma _{2,i}^{2}}}
\le
\sqrt{\sum\nolimits_{i}{\sigma _{2,i}^{2}}}\\
&=&{\left\| {{\rho }_{2}} \right\|_{\operatorname{HS}}}.
\end{eqnarray*}
Hence, we have ${{0 \le {{\mathcal{F}}_{R}}\left( {{\rho }_{1}},{{\rho }_{2}} \right) }} \le 1$ and ${{\Theta }_{R}}=\cos^{-1} {{\mathcal{F}}_{R}}$ is valid. In addition, compared with another recently used fidelity~\cite{QSL10} ${{\mathcal{F}'}}\left( {{\rho }_{1}},{{\rho }_{2}} \right)$\\$={{\left\| \sqrt{{{\rho }_{1}}}\sqrt{{{\rho }_{2}}} \right\|}_{\operatorname{HS}}}/\sqrt{{\left\| {{\rho }_{1}} \right\|}_{\operatorname{HS}} \cdot {\left\| {{\rho }_{2}} \right\|}_{\operatorname{HS}}}$, ${{\mathcal{F}_R}}\left( {{\rho }_{1}},{{\rho }_{2}} \right)$ can guarantee a perfect and simple linear relationship (as we shall see later) at the expense of good symmetry between $\rho_1$ and $\rho_2$.

\subsection*{Minimal evolution time}
The minimal evolution time ${\hat{\tau }}$ of a quantum evolution is defined in the following: given a predefined quantum evolution ${{\dot{\rho }}_{t}}={{\mathcal{L}}_{t}}\left( {{\rho }_{t}} \right)$, then, a predetermined state ${\rho }_{\tau }$, one has ${\hat{\tau }}=\min \{\tau | \tau \in \mathbb{T}\}$, where $\mathbb{T}$ stands for the set of all the actual possible driving time $\tau$ that the evolution from ${\rho }_{0}$ to ${\rho }_{\tau }$ may take. One should notice that $\tau$ is not unique, especially in the non-Markovian strong-coupling regime.

\subsection*{Quantum speed limit}
In order to derive a lower bound as the QSL for driving time $\tau$, the square of the relative-purity fidelity
\begin{equation*}
\label{relative}
{{\left[ {{\mathcal{F}}_{R}}\left( {{\rho }_{t}},{{\rho }_{0}} \right) \right]}^{2}}=\frac{\operatorname{Tr}\left\{ {{\left| \sqrt{{{\rho }_{t}}}\sqrt{{{\rho }_{0}}} \right|}^{2}} \right\}}{\left\| {{\rho }_{0}} \right\|_{\operatorname{HS}}}=\frac{\operatorname{Tr}\left\{ {{\rho }_{0}}{{\rho }_{t}} \right\}}{\left\| {{\rho }_{0}} \right\|_{\operatorname{HS}}}
\end{equation*}
is used, which is simply linear with $\rho_t$. The same linear relationship for $\left[\mathcal{F}_B \left( {{\rho }_{t}},{{\rho }_{0}} \right)\right]^2$ is not true unless $\rho_0$ is a pure state. Taking time derivatives of ${{\Theta }_{R}}=\cos^{-1} {{\mathcal{F}}_{R}}$ yields
\begin{equation*}
\label{inequality2}
{{\dot{\Theta }}_{R}}=\frac{\operatorname{Tr}\left\{ {{\rho }_{0}}{{{\dot{\rho }}}_{t}} \right\}}{-\left\| {{\rho }_{0}} \right\|_{\operatorname{HS}}\sin 2{{\Theta }_{R}}}\le \frac{\left| \operatorname{Tr}\left\{ {{\rho }_{0}}{{\mathcal{L}}_{t}}\left( {{\rho }_{t}} \right) \right\} \right|}{\left\| {{\rho }_{0}} \right\|_{\operatorname{HS}}\sin 2{{\Theta }_{R}}}.
\end{equation*}
The dynamical map of a general quantum system reads ${{\mathcal{L}}_{t}}\left( {{\rho }_{t}} \right)=\left(-i/\hbar\right)[H_{t},{{\rho }_{t}}]+{{\mathcal{D}}_{t}}\left( {{\rho }_{t}} \right)$~\cite{OpenQuantumSystems}, where the renormalized Hamiltonian $H_{t}=H_{0}+H_{t}^{LS}$ contains a time-dependent Lamb shift term $H_{t}^{LS}$. For a Markovian system, the super-operator ${{\mathcal{L}}}$ takes a Lindblad form and is time-independent, hence $\left| \operatorname{Tr}\left\{ {{\rho }_{0}}{{\mathcal{L}}}\left( {{\rho }_{t}} \right) \right\} \right|=\left| \operatorname{Tr}\left\{ \mathcal{L}^{\dagger }\left( {{\rho }_{0}} \right){{\rho }_{t}} \right\} \right|$, where ${\mathcal{L}^{\dagger }}$ obeys the adjoint master equation~\cite{QSLopen1}. However, this is invalid for a non-Markovian system~\cite{OpenQuantumSystems}. To derive the lower bound for a non-Markovian case, we divide ${{\mathcal{L}}_{t}}$ into two parts using the triangle inequality
\begin{equation*}
\label{inequality3}
\left| \operatorname{Tr}\left\{ {{\rho }_{0}}{{\mathcal{L}}_{t}}\left( {{\rho }_{t}} \right) \right\} \right| \le \frac{1}{\hbar }\left| \operatorname{Tr}\left\{ [H_{t},{{\rho }_{t}}]{{\rho }_{0}} \right\} \right|+\left| \operatorname{Tr}\left\{ {{\mathcal{D}}_{t}}\left( {{\rho }_{t}} \right){{\rho }_{0}} \right\} \right|.
\end{equation*}
The absolute trace inequality~\cite{vonNeumann} reads $\left| \operatorname{Tr}\{AB\} \right|\le \min \left\{ \sigma _{1}^{A}  \sum\nolimits_{i}{\sigma _{i}^{B}},\sigma _{1}^{B} \sum\nolimits_{i}{\sigma _{i}^{A}} \right\}.$ Since ${{\left\| {{\rho }_{0}} \right\|}_{\operatorname{tr}}}=1$, one has
\begin{equation}
\label{inequality4}
\left\| {{\rho }_{0}} \right\|_{\operatorname{HS}}{{\dot{\Theta }}_{R}}\sin 2{{\Theta }_{R}}\le \frac{1}{\hbar }{{\left\| [H_{t},{{\rho }_{t}}] \right\|}_{\operatorname{op}}}+{{\left\| {{\mathcal{D}}_{t}}\left( {{\rho }_{t}} \right) \right\|}_{\operatorname{op}}}.
\end{equation}
As $H_{t}$ and ${\rho }_{t}$ are both positive (by shifting the ground energy of $H_{t}$) and Hermitian operators, we take the commutator inequality~\cite{TraceInequality2}, i.e., ${{\left\| AB-BA \right\|}_{\operatorname{op}}}\le \left( \sigma _{1}^{A}-\sigma _{N}^{A} \right)\left( \sigma _{1}^{B}-\sigma _{N}^{B} \right)/2$, where ${{\sigma }_{N}}=\min \left\{ {{\sigma }_{i}} \right\}$ and $N$ is the rank of the operator. For convenience, we denote $\sigma _{1}^{A}-\sigma _{N}^{A}={{\left\| A \right\|}_{\Delta }}$. It is worth noting that this inequality is sharp, e.g., if $A=\left( \begin{matrix}
   2 & 0  \\
   0 & 1  \\
\end{matrix} \right)$ and $B=\left( \begin{matrix}
   1/2 & 1/2  \\
   1/2 & 1/2  \\
\end{matrix} \right)$, then ${{\left\| A \right\|}_{\Delta }}={{\left\| B \right\|}_{\Delta }}=1$ and ${{\left\| AB-BA \right\|}_{\operatorname{op}}}=1/2$. Since ${{\left\| {{\rho }_{t}} \right\|}_{\Delta }}\le 1$, for simplicity we have ${{\left\| H_{t}{{\rho }_{t}}-{{\rho }_{t}}H_{t} \right\|}_{\operatorname{op}}}\le {{\left\| H_{t} \right\|}_{\Delta }}/2$. Substituting it into Eq.~(\ref{inequality4}) and integrating $t$ from $0$ to $\tau$ then yield
\begin{equation}
\label{Final1}
{{\tau }}\ge \frac{\left\| {{\rho }_{0}} \right\|_{\operatorname{HS}}{{\sin }^{2}}{{\Theta }_{R}}}{\frac{1}{2\hbar }{{\left\langle {{\left\| H_{t} \right\|}_{\Delta }} \right\rangle }_{\tau }}+{{\left\langle {{\left\| {{\mathcal{D}}_{t}}\left( {{\rho }_{t}} \right) \right\|}_{\operatorname{op}}} \right\rangle }_{\tau }}},
\end{equation}
where ${{\left\langle A \right\rangle }_{\tau }}={\tau}^{-1}  \int_{0}^{\tau }{Adt}$. It is manifested that Eq.~(\ref{Final1}) is determined by both the renormalized Hamiltonian $H_{t}$ (system) and the dissipator ${{\mathcal{D}}_{t}}\left( {{\rho }_{t}} \right)$ (environment). Also, this bound can reduce to the previous result~\cite{QSLnonmarkov} when ${\rho}_{0}$ is a pure state and $H_{t} \equiv 0$.

\subsection*{Non-Markovianity}
\begin{figure}[h]
	\centerline{\includegraphics[width=8cm]{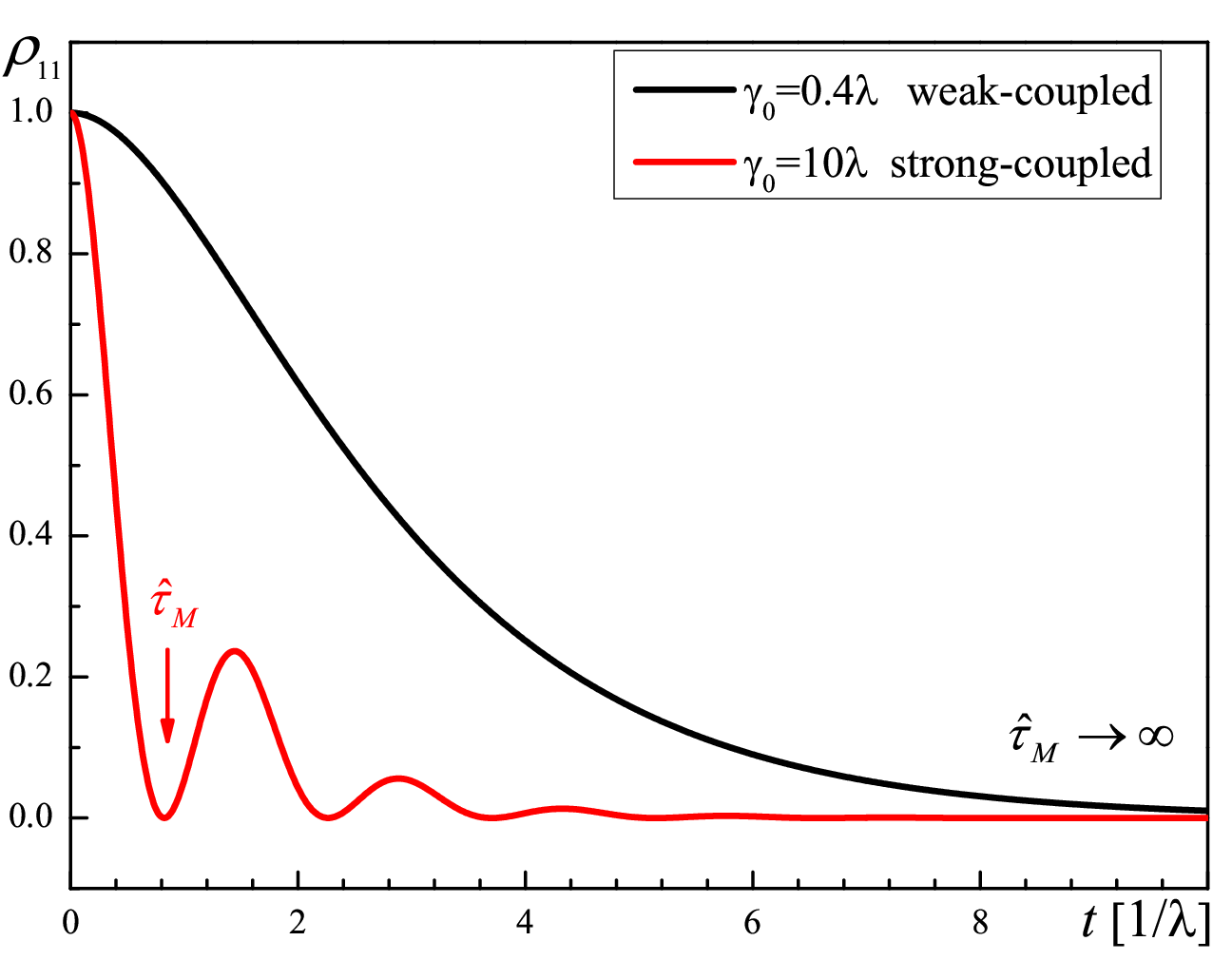}}
	\caption{\label{Fig1}Solutions of the population of the damped Jaynes-Cummings model~\cite{OpenQuantumSystems} in the weak- (black line) and strong-coupling regime (red line), with ${\gamma}_{0}=0.4$ and ${\gamma}_{0}=10$, respectively, and $\lambda =1$ for both. ${\hat{\tau }_{M}}$ is when the maximum of geometric distance is reached (${\rho}_{11}\left(\hat{\tau}_M\right)=0$).\hfill\hfill}
\end{figure}

\begin{figure}[h]
	\centerline{\includegraphics[width=8.5cm]{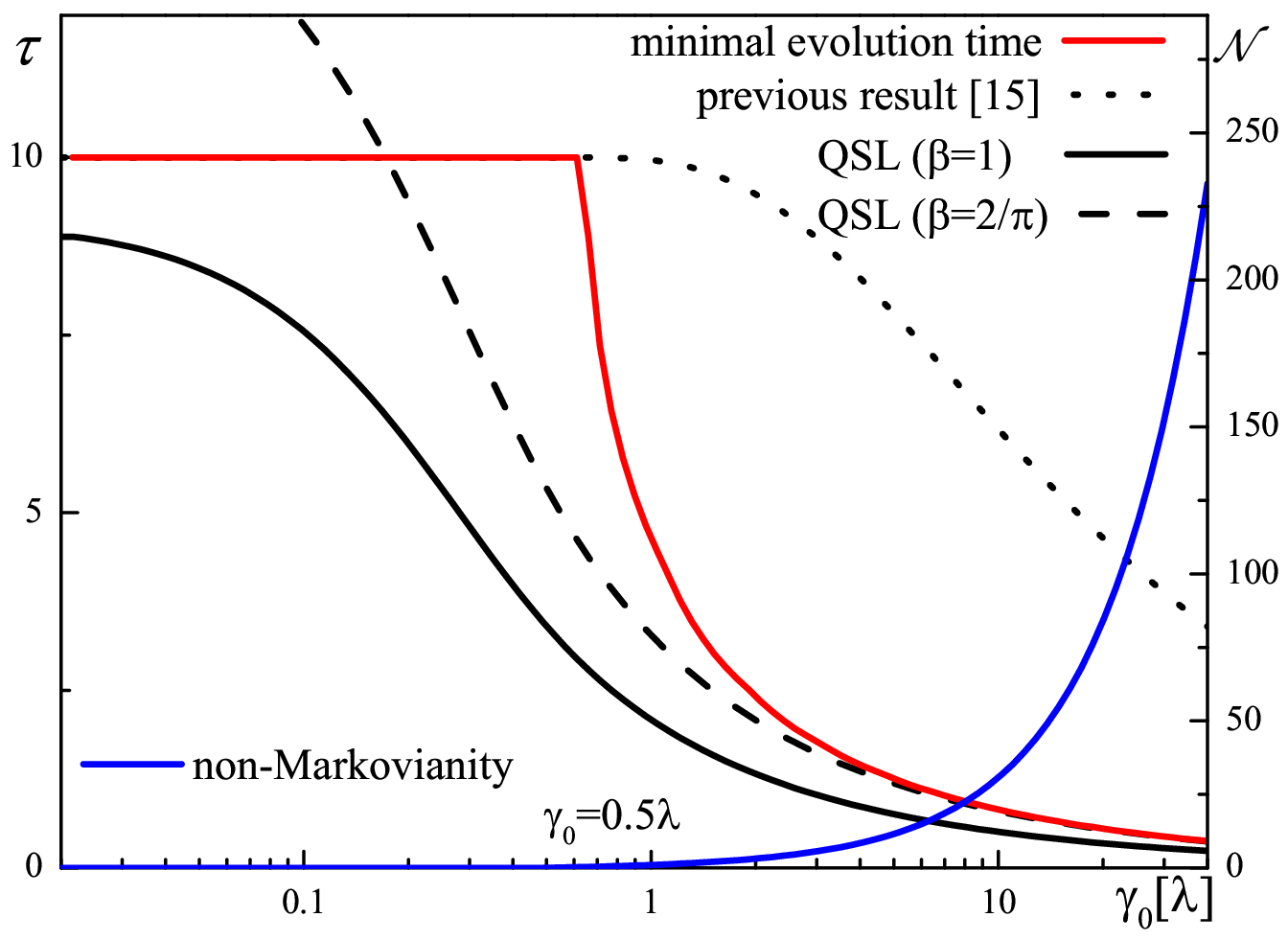}}
	\caption{\label{Fig2}Minimal evolution time (red solid line) of the same model and its different QSL bounds (black lines) as a function of ${\gamma}_{0}$. The bounds are derived from the previous result~\cite{QSLnonmarkov} (dotted), Eq.~(\ref{Final2}) (solid), and Eq.~(\ref{bound2}) (dashed). Also indicated here is the degree of non-Markovianity~\cite{DegreeOfNonMarkovianity1} (blue solid line). We set $\lambda =1$ and $\tau =10$.\hfill\hfill}
\end{figure}

To investigate the minimal evolution time $\hat{\tau}$ in more detail, we use the damped Jaynes-Cummings model as an example, which describes the coupling between a two-level system and a single cavity mode with the background of cavity-QED~\cite{OpenQuantumSystems}. Within a resonant Lorentzian spectral density of environment that $J\left( \omega  \right)=(2\pi)^{-1} {{\gamma }_{0}}{{\lambda }^{2}}/ \left[ {{\left( {{\omega }_{0}}-\omega  \right)}^{2}}+{{\lambda }^{2}} \right]$, the exact Hamiltonians read~\cite{OpenQuantumSystems}
\begin{equation*}
{{H}_{0}}={\hbar{\omega }_{0}}{{\sigma }_{+}}{{\sigma }_{-}}, \qquad H_{t}^{LS}=0,
\end{equation*}
where $\hbar \omega_0$ is the energy difference and ${{\sigma }_{\pm }}=({{\sigma }_{x}\pm i {\sigma }_{y}})/2$ are Pauli operators. The exact dissipator reads
\begin{equation*}
\label{DtJC}
{\mathcal{D}_{t}}\left( {{\rho }_{t}} \right)=\gamma \left( t \right)\left( {{\sigma }_{-}}{{\rho }_{t}}{{\sigma }_{+}}-\frac{1}{2}\{{{\sigma }_{+}}{{\sigma }_{-}},{{\rho }_{t}}\} \right),
\end{equation*}
with $\gamma (t)={2{{\gamma }_{0}}\left( \lambda /D \right)\tan (Dt/2)}/\left[{1+\left( \lambda /D \right)\tan (Dt/2)}\right]$,
in which $\lambda$ is the spectral width, ${\gamma}_{0}$ the coupling strength, and $D=\sqrt{2{{\gamma }_{0}}\lambda -{{\lambda }^{2}}}$. When ${{\gamma }_{0}}<\lambda /2$, the system and environment are weakly coupled and evolve subexponentially; the degree of non-Markovianity $\mathcal{N}=0$~\cite{DegreeOfNonMarkovianity1}. When ${{\gamma }_{0}}>\lambda /2$, $D$ is real; the system and environment are strongly coupled with oscillatory characteristics~\cite{OpenQuantumSystems} and $\mathcal{N}>0$ (see Fig.~\ref{Fig2}). The initial environment is chosen to be a vacuum state and the initial system ${{\rho }_{0}}=\left( \begin{matrix}
   1 & 0  \\
   0 & 0  \\
\end{matrix} \right)$ fully excited to make the model simpler. Consequently, we only need to consider the dissipator ${{\mathcal{D}}_{t}}\left( {{\rho }_{t}} \right)$, for the exact solution~\cite{OpenQuantumSystems} implies ${\rho }_{t}$ a diagonal operator so that $[H_{t},{{\rho }_{t}}]\equiv0$ in Eq.~(\ref{inequality4}). It is useful to introduce a special minimal evolution time ${\hat{\tau }_{M}}=\max \left\{ \hat{\tau}  \right\}$, $\forall \mathbb{T}$, i.e., ${\hat{\tau }_{M}}$ is the minimal evolution time for the maximum of ${{\Theta }_{R}}$. It is worth noting that ${\hat{\tau }_{M}}$ depends strongly on different coupling regimes (see Fig.~\ref{Fig1}): in the weak-coupling regime, we have ${\hat{\tau }_{M}}\to \infty $, but in the strong-coupling case, ${\hat{\tau }_{M}}$ is finite, which is caused by the oscillatory characteristics of the population. Like ${\hat{\tau }_{M}}$, it is worth noting for ${\hat{\tau }}$ itself that it will be smaller in the strong-coupling regime than that in the weak-coupling regime~\cite{QSLnonmarkov}. Although ${\hat{\tau }}$ is equal to the only possible driving time $\tau \in \mathbb{T}$ when it is weakly coupled, it is not the case for the strong-coupling regime. From numerical solution of ${\hat{\tau }}$ we find that ${\hat{\tau }}$ has a first derivative singular point at ${{\gamma }_{0}} \approx \lambda /2$ and a steep decrease in the strong-coupling regime, which cannot be implied from $\mathcal{N}$ (see Fig.~\ref{Fig2}). A decrease of the QSL for ${\hat{\tau }}$ was also suggested in the previous result~\cite{QSLnonmarkov}, but the decreasing slope with ${\gamma}_{0}$ deviates from the minimal evolution time as shown in Fig.~\ref{Fig2}.


As the energy of an open system is not conserved, the average of the dissipator ${{\mathcal{D}}_{t}}\left( {{\rho }_{t}} \right)$ decreases with time and ${{\left\langle {{\left\| {{\mathcal{D}}_{t}}\left( {{\rho }_{t}} \right) \right\|}_{\operatorname{op}}} \right\rangle }_{\infty }}=0$; as a result, we have ${{\left\langle {{\left\| {{\mathcal{D}}_{t}}\left( {{\rho }_{t}} \right) \right\|}_{\operatorname{op}}} \right\rangle }_{\tau }} \le {{\left\langle {{\left\| {{\mathcal{D}}_{t}}\left( {{\rho }_{t}} \right) \right\|}_{\operatorname{op}}} \right\rangle }_{\hat{\tau}}}$ since $\tau \ge \hat{\tau}$. $\hat{\tau}$ depends on the short duration from $0$ to $\hat{\tau }_{M}$ at most, so we can simply replace the time average by the maximum and eliminate the subscript $\hat{\tau}$,
\begin{equation}
\label{bound1}
{{\left\langle {{\left\| {{\mathcal{D}}_{t}}\left( {{\rho }_{t}} \right) \right\|}_{\operatorname{op}}} \right\rangle }_{\hat{\tau}}} \le \max \left\{ {{\left\| {{\mathcal{D}}_{t}}\left( {{\rho }_{t}} \right) \right\|}_{\operatorname{op}}} \right\}.
\end{equation}
Substituting Eq.~(\ref{bound1}) into Eq.~(\ref{Final1}), the final bound for $\hat{\tau}$ yields
\begin{equation}
\label{Final2}
\hat{\tau} \ge \frac{\left\| {{\rho }_{0}} \right\|_{\operatorname{HS}}{{\sin }^{2}}{{\Theta }_{R}}}{\frac{1}{2\hbar }{{\left\langle {{\left\| H_{t} \right\|}_{\Delta }} \right\rangle }_{\hat{\tau }}}+\max \left\{ {{\left\| {{\mathcal{D}}_{t}}\left( {{\rho }_{t}} \right) \right\|}_{\operatorname{op}}} \right\}},
\end{equation}
which is valid for general quantum systems, regardless of whether they are closed or open and how strong the coupling is. It is found that the QSL Eq.~(\ref{Final2}) in the strong-coupling regime has a fitting decreasing slope as shown in Fig.~\ref{Fig2}. However, this bound is not asymptotic when ${{\gamma }_{0}}/\lambda \to \infty$. To derive a sharper QSL, we notice that Eq.~(\ref{bound1}) can take an approximation,
\begin{equation}
\label{bound2}
{{\left\langle {{\left\| {{\mathcal{D}}_{t}}\left( {{\rho }_{t}} \right) \right\|}_{\operatorname{op}}} \right\rangle }_{\hat{\tau}}}\approx \beta \max \left\{ {{\left\| {{\mathcal{D}}_{t}}\left( {{\rho }_{t}} \right) \right\|}_{\operatorname{op}}} \right\},
\end{equation}
where the parameter $\beta$ introduced as a metric of the time average rests upon specific models, and the rough bound of Eq.~(\ref{bound1}) can also be treated as $\beta=1$. For this case, we consider that in the strong-coupling regime when $\lambda /D\to 0$,
\begin{equation*}
\label{gamma1}
\frac{\gamma (t)}{2{{\gamma }_{0}}}=\frac{\lambda }{D}\tan \left(\frac{Dt}{2}\right)-{{\left[ \frac{\lambda }{D}\tan \left(\frac{Dt}{2}\right) \right]}^{2}}+\cdots.
\end{equation*}
The first-order approximation yields $\gamma (t)\approx D\tan (Dt/2)$, since $2{{\gamma }_{0}}\lambda \approx {{D}^{2}}$. The exact solution of ${{\dot{\rho }}_{t}}={{\mathcal{D}}_{t}}\left( {{\rho }_{t}} \right)$ yields ${{\mathcal{D}}_{t}}\left( {{\rho }_{t}} \right)={{\mathcal{D}}_{t}}\left[ {{\mathcal{T}}_{\leftarrow }}\exp \left( \int_{0}^{t}{d{t'}{{\mathcal{D}}_{{{t'}}}}} \right) \right]{{\rho }_{0}}$. Here ${{\mathcal{T}}_{\leftarrow }}$ is the chronological super-operator which orders the $t'$ arguments to increase from right to left~\cite{TOperator}. Hence,
\begin{equation}
\label{oscillator}
{{\left\| {{\mathcal{D}}_{t}}\left( {{\rho }_{t}} \right) \right\|}_{\operatorname{op}}}
\approx
\max \left\{ {{\left\| {{\mathcal{D}}_{t}}\left( {{\rho }_{t}} \right) \right\|}_{\operatorname{op}}} \right\}
\left| \sin \left( Dt \right) \right|,
\end{equation}
where $\max \left\{ {{\left\| {{\mathcal{D}}_{t}}\left( {{\rho }_{t}} \right) \right\|}_{\operatorname{op}}} \right\} \approx D/2$. Generally speaking, for a continuous evolution, strong coupling between the system and environment certainly involves a non-Markovian bidirectional flow of energy and/or coherence, which can always be characterized as oscillator(s). Therefore, a general form like Eq.~(\ref{oscillator}) can provide a reasonable approximation of oscillation for other models.

It is found that the parameter $\beta$ depends typically on the relation between ${\hat{\tau}}$ and ${{\hat{\tau}}_{M}}$ in the strong-coupling limit. With different ${\hat{\tau}}/{{\hat{\tau}}_{M}}$ (when ${{\gamma }_{0}} /\lambda \to \infty$), the time average in the left term of Eq.~(\ref{bound2}) will take a different time period, and $\beta$ thus changes in the range from $0$ to $1$. In this case of the damped Jaynes-Cummings model, we have ${\hat{\tau}}/{{\hat{\tau}}_{M}}\to 1$ as ${{\gamma }_{0}} /\lambda \to \infty$, suggesting that the time average in Eq.~(\ref{bound2}) should take nearly a $\pi/2$ period. Taking Eq.~(\ref{oscillator}) into Eq.~(\ref{bound2}) immediately indicates $\beta=2/\pi$ then. From Fig.~\ref{Fig2}, it is clear that this bound is sharp, but is not valid when it comes into the weak-coupling regime since the approximation $\gamma (t)\approx D\tan (Dt/2)$ is invalid there.

\subsection*{Renormalized Hamiltonian}
\begin{figure}[h]
	\centerline{\includegraphics[width=8.5cm]{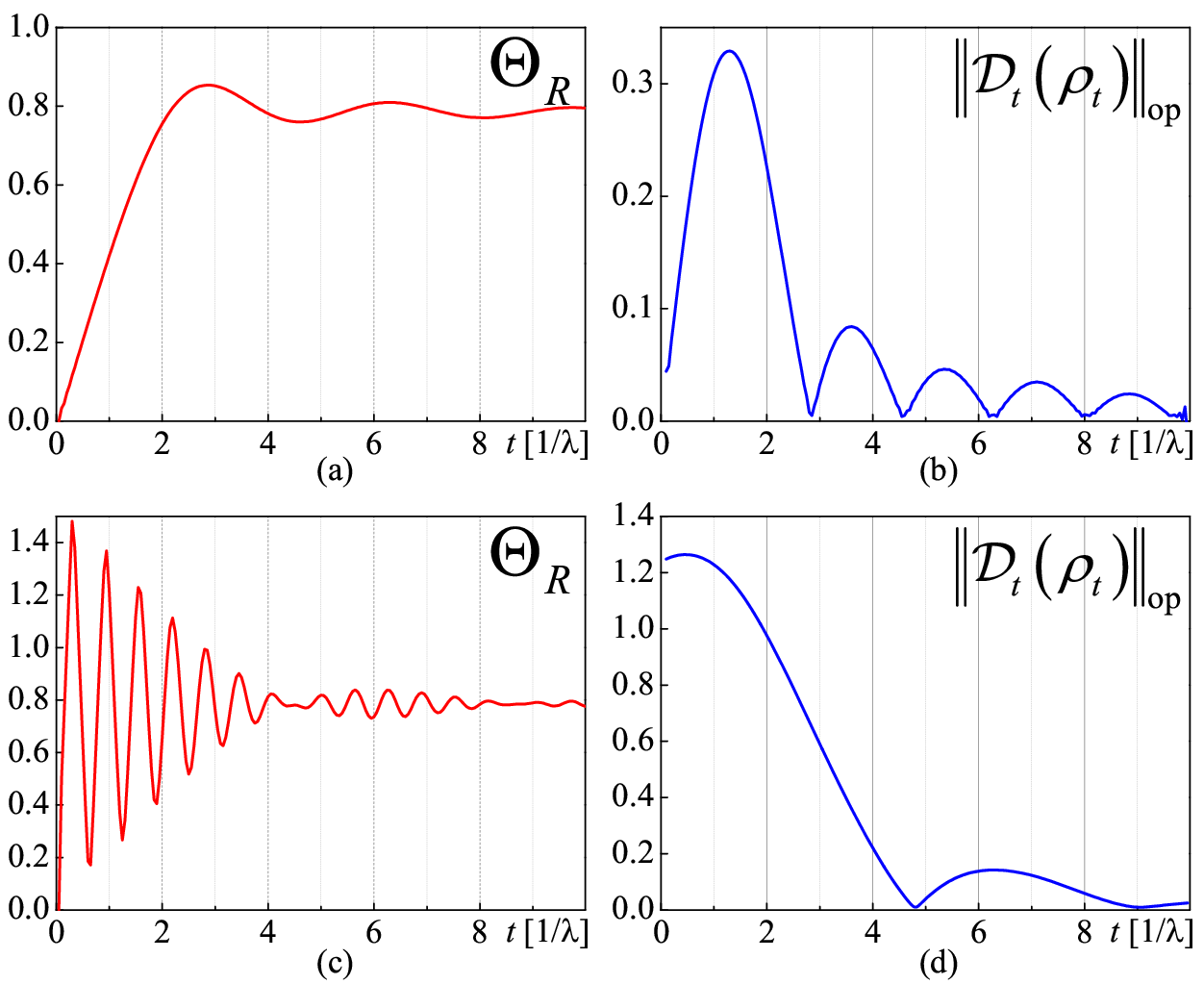}}
	\caption{\label{Fig3}Numerical solution of the relative-purity fidelity (red lines) and the dissipator (blue lines) of the quantum dot model~\cite{TwoBandsModel2}. $\lambda =0.02$ which represents the strong-coupling regime, with ${N}_{1}={N}_{2}=500$ and $\delta \varepsilon=0.5\hbar$. The initial states are (a),~(b):~$\left( \protect\begin{matrix}
		1 & 0  \\
		0 & 0
		\protect\end{matrix} \right)$  and (c),~(d):~$\left( \protect\begin{matrix}
		1/2 & 1/2  \\
		1/2 & 1/2
		\protect\end{matrix} \right)$, respectively.\hfill\hfill}
\end{figure}

To verify our result and manifest the influence of the renormalized Hamiltonian term in Eq.~(\ref{Final1}), we introduce another two-level system containing a two-band environment as the second example. This model can simulate the interaction between a spin and a single-particle quantum dot~\cite{TwoBandsModel2,NMQT}, of which the total Hamiltonian is $H={H}_{0}+V$ where ${{H}_{0}}=\Delta E{{\sigma }_{z}}+\sum\nolimits_{{{n}_{1}}}{\left({\delta \varepsilon }/{{{N}_{1}}}\right)}{{n}_{1}}\left| {{n}_{1}} \rangle\langle {{n}_{1}}\right|+\sum\nolimits_{{{n}_{2}}}{\left( \Delta E+{\delta \varepsilon }/{{{N}_{2}}} \right)}{{n}_{2}}\left| {{n}_{2}} \rangle\langle  {{n}_{2}} \right|$ with ${\sigma }_{z}$ the Pauli operator. The lower energy band contains ${N}_{1}$ levels and the upper ${N}_{2}$ levels, with the same band width $\delta \varepsilon$ and the inter-bands distance $\Delta E$ in resonance with the spin. $V$ represents the interaction that $V=\lambda \sum\nolimits_{{{n}_{1},n_2}}{c(n_1,n_2) \sigma_{+}} \left| {{n}_{1}} \rangle\langle {{n}_{2}}\right|+\text{h.c.}$, with $\lambda$ the coupling coefficient and $c(n_1,n_2)$ complex Gaussian random variables. At the beginning, we numerically solve the model concerning the minimal evolution time problem and identify the same singularity at $\lambda \approx 0.0072$ and steep decrease when $\lambda > 0.0072$ like those shown in Fig.~\ref{Fig2}. To demonstrate the influence of renormalized Hamiltonian, first we set ${{\left( \rho \otimes {{\rho }_{E}} \right)}_{0}}=\left( \begin{matrix}
   1 & 0  \\
   0 & 0  \\
\end{matrix} \right) \otimes \left| {{n}_{1}} \rangle\langle  {{n}_{1}} \right|$ with a driving time $\tau =8.0$, from which one derives ${{\Theta }_{R}}\approx 0.7707$ and ${\hat{\tau}}\approx 2.0$ (see Figs.~\ref{Fig3}(a),~(b)). As ${{\rho }_{\infty }}=\left( \begin{matrix}
   1/2 & 0  \\
   0 & 1/2  \\
\end{matrix} \right)$, ${\hat{\tau}}/{{\hat{\tau}}_{M}} \nrightarrow 1$ now~\cite{TwoBandsModel2}. It is recalculated from Eq.~(\ref{bound2}) that $\beta \approx \left[ 1-\cos \left( 3 \pi /4 \right) \right]/\left( 3\pi /4 \right)\approx 0.72$. Further calculation shows that the previous QSL~\cite{QSLnonmarkov} $\hat{\tau}_{o}^{\text{QSL}}=5.1757$ is too large, while Eqs.~(\ref{bound1})~and~(\ref{bound2}) indicate $\hat{\tau}_{\beta =1}^{\text{QSL}}=1.4421$ and $\hat{\tau}_{\beta =0.72}^{\text{QSL}}=1.9905$. Both of them stay valid while the latter is sharp. Second, we set $\Delta E=10\hbar$, ${{\left( \rho \otimes {{\rho }_{E}} \right)}_{0}}=\left( \begin{matrix}
   1/2 & 1/2  \\
   1/2 & 1/2  \\
\end{matrix} \right) \otimes \left| {{n}_{1}} \rangle\langle  {{n}_{1}} \right|$ and $\tau=8.0$, from which one derives ${{\Theta }_{R}}\approx 0.7832$ and $\hat{\tau }\approx 0.2$ (see Figs.~\ref{Fig3}(c),~(d)). Since $\rho_0$ is not diagonal, $[H_{t},{{\rho }_{t}}]\ne 0$ and ${{\left\langle {{\left\| H_{t} \right\|}_{\Delta }} \right\rangle }_{\tau }}$ should be considered. Further calculation shows that the previous QSL~\cite{QSLnonmarkov} $\hat{\tau}_{o}^{\text{QSL}}=1.0242$ is too large, while $\hat{\tau}_{\beta =1}^{\text{QSL}}=0.1130$ and $\hat{\tau}_{\beta =0.72}^{\text{QSL}}=0.1196$. The mere difference between $\hat{\tau}_{\beta =1}^{\text{QSL}}$ and $\hat{\tau}_{\beta =0.72}^{\text{QSL}}$ implies that the renormalized Hamiltonian $H_{t}$ is dominant in Eq.~(\ref{Final2}). As ${{\left\langle {{\left\| H_{t} \right\|}_{\Delta }} \right\rangle }_{\tau }}$ added, ${\hat{\tau}^{\text{QSL}}}$ becomes smaller, which apparently follows the time-energy uncertainty relation.

\section*{Discussion}
Only Hamiltonian was considered in some of previous investigations~\cite{MT,ML,QSL8,QSL9,QSL5,QSL3,QSL2}, while for an open system, the coupling strength of its dissipator also has an influence on QSL~\cite{QSLopen1, QSLnonmarkov}. However, it is demonstrated in our study that in non-Markovian case such influence could be more significant than it was thought. Therefore, to achieve a high speed of evolution~\cite{QuantumSpeedUp}, it is more probable that we only focus on improving the coupling interaction instead of increasing the energy. This implies that the power consumption can stay a low level for cavity-QED process while high efficiency can still be achieved. Previously it was always thought that a strong coupling with environment should be prevented due to its enhanced decoherence effect on qubits. However, as a trade-off, the operation time for transforming and/or erasing qubits for example can also be remarkably reduced in the strong-coupling regime. It is thus possible to make quantum computation more feasible and achievable by adjusting the coupling strength in a well-chosen pattern.

In summary, we derive a sharp bound as the quantum speed limit of open systems available for mixed initial states. Considering the non-Markovian feature, we find that the minimal evolution time of the two two-level examples considered here has singularity nearly at the cross-point of regimes and a steep decrease in the strong-coupling regime. This result may lead to high-efficiency quantum information research and engineering. As the time-energy uncertainty relation dictates, renormalized Hamiltonian will also contribute to the final quantum speed limit bound as manifested in the quantum dot model in detail. We expect our result to be used for quantum time analysis and optimal control, as well as in pertinent topics on general physics.

\section*{Methods}
\subsection*{Norms of operators}
A general Schatten $p$-norm of an operator $A$ is ${{\left\| A \right\|}_{p}}={{\left( \sum\nolimits_{i}{\sigma _{i}^{p}} \right)}^{1/p}}$, where singular values ${{\sigma }_{1}}\ge {{\sigma }_{2}}\ge \ldots \ge {{\sigma }_{i}}\ge \ldots \ge 0$ are the eigenvalues of $\left| A \right|=\sqrt{{{A}^{\dagger }}A}$, and ${{\left\| A \right\|}_{\operatorname{op}}}={{\left\| A \right\|}_{p\to \infty }}={{\sigma }_{1}}$, ${{\left\| A \right\|}_{\operatorname{tr}}}={{\left\| A \right\|}_{p=1}}$ and ${{\left\| A \right\|}_{\operatorname{HS}}}={{\left\| A \right\|}_{p=2}}$ as the operator norm, trace norm and Hilbert-Schmidt norm of $A$, respectively~\cite{Norms2}.

\subsection*{Approximation for the dissipator ${{\left\| {{\mathcal{D}}_{t}}\left( {{\rho }_{t}} \right) \right\|}_{\operatorname{op}}}$ of the damped Jaynes-Cummings model in the strong-coupling regime}
With
\begin{equation*}
{\mathcal{D}_{t}}\left( {{\rho }_{t}} \right)=\gamma \left( t \right)\left( {{\sigma }_{-}}{{\rho }_{t}}{{\sigma }_{+}}-\frac{1}{2}\{{{\sigma }_{+}}{{\sigma }_{-}},{{\rho }_{t}}\} \right)
\end{equation*}
given~\cite{OpenQuantumSystems}, the exact solution of ${{\dot{\rho }}_{t}}={{\mathcal{D}}_{t}}\left( {{\rho }_{t}} \right)$ yields
\begin{eqnarray*}
{{\mathcal{D}}_{t}}\left( {{\rho }_{t}} \right)&=&{{\mathcal{D}}_{t}}\left[ {{\mathcal{T}}_{\leftarrow }}\exp \left( \int_{0}^{t}{d{t'}{{\mathcal{D}}_{{{t'}}}}} \right) \right]{{\rho }_{0}}\nonumber\\
&=& -\gamma \left( t \right){{\mathcal{T}}_{\leftarrow }}\exp \left( -\int_{0}^{t}{d{t}'\gamma \left( {{t}'} \right)} \right) \left( \begin{matrix}
   1 & 0  \\
   0 & 0  \\
\end{matrix} \right).
\end{eqnarray*}
We have $\gamma (t)\approx D\tan (Dt/2)$ in the strong-coupling regime. As a result,
\begin{eqnarray}
\label{oscillatorb1}
&&{{\left\| {{\mathcal{D}}_{t}}\left( {{\rho }_{t}} \right) \right\|}_{\operatorname{op}}}\nonumber\\
&\approx&\left| D\tan \left( \frac{Dt}{2} \right)\exp \left[ -D\int_{0}^{t}{\tan \left( \frac{Dt'}{2} \right)dt'} \right] \right|\nonumber\\
&=&\frac{D}{2}\left| {{\sin }}\left( {Dt} \right) \right| \propto \max \left\{ {{\left\| {{\mathcal{D}}_{t}}\left( {{\rho }_{t}} \right) \right\|}_{\operatorname{op}}} \right\}\left| \sin \left( Dt \right) \right|,\nonumber
\end{eqnarray}
which yields the result of Eq.~(\ref{oscillator}). In this case, we also have ${\hat{\tau }}\to {\hat{\tau }_{M}}$ as ${\gamma}_{0}$ increases, which implies that the time average in Eq.~(\ref{bound2}) takes nearly a $\pi/2$ period. Taking Eq.~(\ref{oscillator}) into Eq.~(\ref{bound2}) then indicates $\beta=2/\pi$.

\bibliography{Bib}

\section*{Acknowledgements}
We thank Jian-Wei Zhang, Yusui Chen, Paul Berman, and Yang Li for their valuable advice. This work is supported by the National Science Fund for Distinguished Young Scholars of China (Grant No. 61225003), National Natural Science Foundation of China (Grant No. 61101081), and the National Hi-Tech Research and Development (863) Program.

\section*{Author contributions}
X.M., C.W., and H.G. calculated and analyzed the results. X.M. and H.G.
co-wrote the paper. All authors reviewed the manuscript and agreed with the submission.

\section*{Additional information}
\textbf{Competing financial interests:} The authors declare no competing financial interests.

\end{document}